\documentclass[%
 reprint,
 amsmath,amssymb,
 aps,
prd,
]{revtex4-2}

\usepackage{graphicx}
\usepackage{dcolumn}
\usepackage{bm}
\usepackage[usenames, dvipsnames]{color}

\begin{document}

\author{Kiara C. Ruffin}
\affiliation{University of Virginia, Charlottesville, Virginia 22904, USA}
\email[]{kcr9jqr@virginia.edu}

\title{Parametric Resonance in RF Axion Haloscopes}
\author{Gray Rybka}
\affiliation{University of Washington, Seattle, Washington 98195, USA}
\email[]{grybka@uw.edu}

\begin{abstract}
The axion were proposed as a result to a solution to the Strong CP Problem in quantum chromodynamics (QCD) and is now considered a leading candidate for dark matter.  Direct axion dark matter detection experiments are challenging due to the axion's weak interaction with electromagnetism.  Recent work has suggested the possibility of an enhancement of astrophysical axion-to-photon decay through parametric resonance.  We explore here the feasibility of using parametric resonance to enhance the signal in direct axion-like particle dark matter detectors.
\end{abstract}

\maketitle

\section{Introduction}

New light pseudoscalar particles are viable candidates for dark matter. Of these, the QCD axion is notably well motivated\cite{Preskill:1982cy,Abbott:1982af,Dine:1982ah}, having been first introduced as a solution to symmetry violation in the Strong Charge-Parity (CP) problem in QCD \cite{Peccei+1977,Weinberg1978, Wilczek1978}, but more general axion-like particles are also a possibility \cite{Svrcek_2006}.  Efforts to detect the signatures of axion-like dark matter, both astrophysically and in terrestrial laboratories are underway. In both cases, the axion-to-photon coupling plays a significant role. To persist as dark matter, the lifetime ALPs must be much longer than the age of the universe, but ALP-to-photon decay may be stimulated in the presence of electromagnetic fields. Among laboratory experiments, most sensitive experiment class is the axion haloscope \cite{PhysRevLett.52.695.2}, which stimulates axion decay to photons with a strong static magnetic field, and further enhances the decay rate with an electromagnetic resonance.  Similar effects have been proposed for astrophysical signatures in both static magnetic fields \cite{Raffelt:1996wa} and oscillating fields \cite{PhysRevLett.123.131804,ghosh2020axiongegenscheinprobingbackscattering}.


While astronomical observations can estimate the distribution of dark matter on galactic scales, the small-scale structure is yet to be determined.  The simplest model is that dark matter is smoothly distributed on the solar-system scale \cite{PhysRevD.42.3572}, but much work has suggested that axion dark matter may have substructure in the form of clumps or miniclusters \cite{HOGAN1988228} with a variety of masses depending on details of axion self-interactions and cluster formation history \cite{Kolb_1996,PhysRevD.106.103514,PhysRevD.101.083014}.  In all cases the density of dark matter axions will be enhanced at the center of the minicluster, offering an opportunity for a stronger axion-to-photon conversion signal.

It has been suggested that a parametrically resonant enhancement of axion-to-photon conversion in axion miniclusters \cite{Hertzberg_2018} could produce a detectable astrophysical signal.  In this case, axion decay into photons of one propogating mode stimulates further decay into another mode, which then in turn enhances the first.  For appropriate values, it was shown that this is the phenomenon of parametric resonance and could significantly enhance axion-to-photon conversion rate. We explore here whether a similar effect could be used in a terrestrial experiment.

\section{The Axion RF Haloscope}

The combination of modes used in \cite{Hertzberg_2018} suggests utilizing an RF axion haloscope as discussed in \cite{sikivie2013superconducting} can be helpful in axion detection.  In this design, instead of a static magnetic field, the excitation of one resonant mode stimulates axion decay into another mode with a frequency difference equal to the axion mass ($m_a \approx \hbar \omega$). 

To illustrate this, we begin with Maxwell's Equations modified to include the coupling to an axion field, where terms involving $\nabla a$ are set to zero because dark matter is expected to be nonrelativistic, following Refs. 
\cite{PhysRevLett.51.1415,PhysRevLett.58.1799,PhysRevD.99.055010}.
\begin{align}
\nabla \cdot E &= \rho \\
\nabla \cdot B &= 0 \\
\nabla \times E &= - \dot{B} \\
\nabla \times B &= \dot{E} + g \dot{a} B.
\end{align}

Here $E$ is the electric field, $B$ is the magnetic field, $J$ is electric current density, $g$ is the axion photon coupling, and $a$ is the magnitude of the axion field.  Inside a resonator, $\rho$ and $J$ are zero except at the boundaries - where currents can be treated as losses or driving terms as appropriate.

The number density of axions is assumed to be large due to the density of dark matter, and as dark matter is non-relativistic, the wavelength of the axion field is assumed to be much larger than the scale of experiments currently being conducted.  Therefore the axion field can be treated as a classical field \cite{cheong2024quantumdescriptionwavedark} oscillating at a frequency $\omega_a=\frac{m_a c^2}{\hbar}$ and can be written as
\begin{equation}
a=a_0 \sin \omega_a t,
\end{equation}
with the relationship between the axion field magnitude and the dark matter density being $a_0^2m_a^2=\rho_{\mathrm{dm}}$.

From these we can develop the wave equation for a bound resonant electromagnetic mode with a resonant frequency $\omega_i$, a loss at the boundaries, and signal readout parameterized by $\zeta_i$ and allowing for driving terms from antennas:
\begin{equation}
-\omega_i^2 E - \ddot{E}-2\Gamma_i\dot{E} = g \ddot{a} B + g \dot{a}\dot{B} +\textrm{driving terms} \label{eqn:waveqn}
\end{equation}

The axion field couples modes with non-orthogonal $E$ and $B$ components. The derivation of this coupling is further addressed in \cite{Berlin2020} showing that if a RF signal is injected into a single mode, the nonlinear axion coupling can cause an excitation of another mode with a frequency at the sum or difference of the injected frequency and the axion frequency.  The effect is proportional to the overlap between the electric and magnetic fields of the two modes, 
\begin{equation}
\eta=\frac{ \int_V E_1^* \cdot B_2}{\sqrt{\int_V |E_1|^2}\sqrt{\int_V |E_2|^2}}.
\label{eqn:wave}
\end{equation}

Modes with large $\eta$ are well-coupled by the axion field.  We will follow a similar procedure, but focusing on the parametric resonant case.

\section{Parametric Resonance in an Axion RF Haloscope}

The axion RF haloscope can be interpreted as a parametric amplifier with the axion field acting as the pump, where there is injected RF power at the `idler' frequency, and the output is measured at a `signal' frequency $\pm \omega_a$.  From a particle physics perspective, the injected RF power stimulates axions to decay to a photon with frequency $\pm\omega_a$ from the injected power.  A parametric amplifier with a strong enough pump will act as a spontaneously radiating parametric resonator.  In this case, the idler stimulates the pump to decay to the signal frequency, and the signal stimulates the pump to decay to the idler frequency.  If the rate of stimulated axion decay is larger than the rate at which photons are lost, the system is unstable and the photon density at the signal and idler frequencies will exponentially grow until limited by other factors.  Therefore, we seek to find the conditions under which an axion RF haloscope can minimize losses and maximize coupling to axions to enter the unstable regime.  

\subsection{Single Mode RF Haloscopes}

A classical example of parametric resonance is the child on a swing. The swing with experiences frequency $\omega_0$ as the child drives the amplitude through pumping their legs producing a frequency $2\omega_0$.  In general, parametric resonance allows energy transfer between two resonant modes and an excitation at the sum or difference frequency of those two modes.

 To replicate the simple example this behavior in an RF haloscope, we consider an RF haloscope's modes as described above that are degenerate at frequency $\omega_0=\frac{\omega_a}{2}$.  Standard haloscopes with cavities based on a right circular cylinder do not typically have this feature.  It has been shown, however that twisted ``chiral'' cavities do have modes degenerate modes with a non-zero $\eta$ parameter, and this can be leveraged for axion searches \cite{PhysRevD.108.052014} (the $\eta$ parameter in this reference is called the helicity).  The degenerate modes in cavity arrangement are particularly convenient for demonstrating parametric resonance.

\subsection{Parametric Resonance Solution}

We suppose that we have a resonant mode described above with a nonzero $\eta$, and a resonant frequency at $\omega_0=\omega_a/2$, and can rewrite eqn.~\ref{eqn:wave} as
\begin{align}
\ddot{E}+\left(2\Gamma-2g\eta a_0 \omega_0 \sin 2\omega_0 t\right)\dot{E}&\\
+\omega_0^2\left(1-4g\eta a_0 \cos 2 \omega_0 t\right)E&=0. \label{eqn:eom}
\end{align}

\begin{figure}
\includegraphics[width=6.5cm]{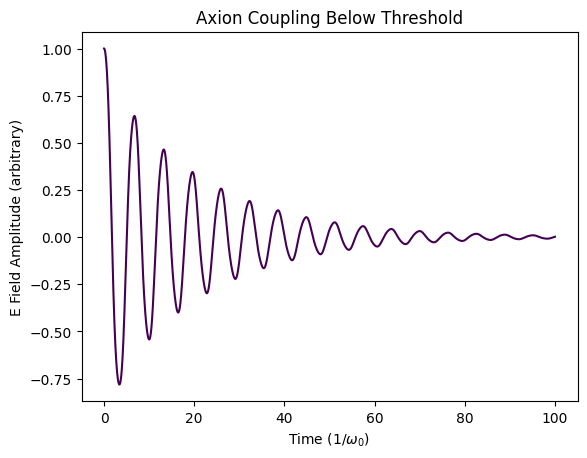}
\includegraphics[width=6.5cm]{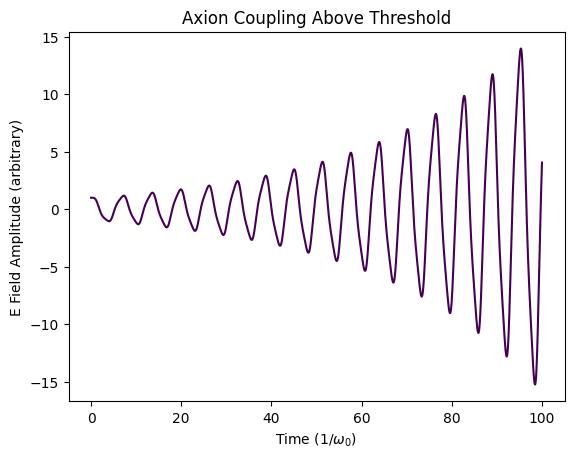}
\caption{Simulations of the electric field in a resonant system described in the presence of an axion field (Eqn.~\ref{eqn:eom}).  The parameters used to generate the top plot are below the instability condition in Eqn.~\ref{eqn:resonance_condition} and the field shows exponential decay.  The parameters used to generate the bottom plot are above the instability threshold and exhibit exponential growth.}
\label{fig:sims}
\end{figure}

This system manifests unstable parametric resonance for a large enough couplings and densities, and shown in Fig.~\ref{fig:sims}.  It can be quantified by noting that in the expected solution $\dot{a}\dot{E}\simeq-\frac{1}{2}\ddot{a}E$, and thus this can can be approximated and re-written in the form of the Mathieu equation with damping \cite{10.1115/1.4039144}. 


\begin{equation}
\ddot{E}+2\Gamma\dot{E}+\omega_0^2\left(1-2g\eta a_0\cos 2 \omega_0 t\right) E\simeq 0.
\end{equation}  

For this equation, we expect unstable (exponentially growing) parametric resonant solutions for $E$ when

\begin{equation}
g\eta a_0 > 2 \frac{\Gamma_i}{\omega_0}.
\label{eqn:resonance_condition}
\end{equation}

In this case the amplitude of the electric field in the cavity will grow as

\begin{equation}
E(t)\propto \exp{ \left( \frac{1}{2}g\eta a_0 -\Gamma_i\right)t}.
\end{equation}

Thus the conclusion is that parametric oscillation can be achieved for large enough axion density and axion-photon coupling, and small enough mode losses.  Interestingly, resonator volume is irrelevant.

\section{Feasibility}

We consider the question of what technical advances would be necessary to achieve an unstable parametric resonant solution with plausible numerical values.  We take the density of axions to be $\rho_a=a_0^2 \omega_a^2$ with a characteristic scale of $\rho_a$ at Earth being 0.45 GeV/cc \cite{OHare:20246n}.  However, in many models axions form femto-clusters~\cite{Kolb_1996}, mini-clusters~\cite{PhysRevD.101.083014}, and even axion stars~\cite{Witte_2023} or clouds around neutron stars~\cite{PhysRevX.14.041015} with densities many orders of magnitude larger than the background dark matter density at their center.  The predicted axion photon coupling for the QCD axion is related to the axion mass, but experimentally the coupling for axion like particles is bounded to be $g_{a\gamma\gamma}< 10^{-11}~\mathrm{GeV^{-1}}$ \cite{Anastassopoulos2017} except in regions already covered by haloscopes, and certain regions excluded by galactic transparency.  Chiral cavity modes can be made with $\eta\simeq\mathcal{O}(1)$, with the frequency set by the cavity length scale.  Cavity losses $\Gamma_i$ are traditionally expressed by resonant quality factor $Q=\omega_0/\Gamma_i$.  Resonant cavity $Q$s have been demonstrated in the microwave regime up to $10^{11}$ \cite{PhysRevApplied.13.034032} and used in axion-like particle searches \cite{PhysRevD.110.043022} .  We take for granted that the axion mass is known and the cavity resonant is ideally tuned, while recognizing that any detuning could be treated as a lowered $Q$.  With these definitions, Eqn.~\ref{eqn:resonance_condition} becomes

\begin{equation}
    g\eta Q\sqrt{\rho_a} > 2 \omega_a.
\end{equation}

\begin{figure}
\centering
\includegraphics[width=6.5cm=]{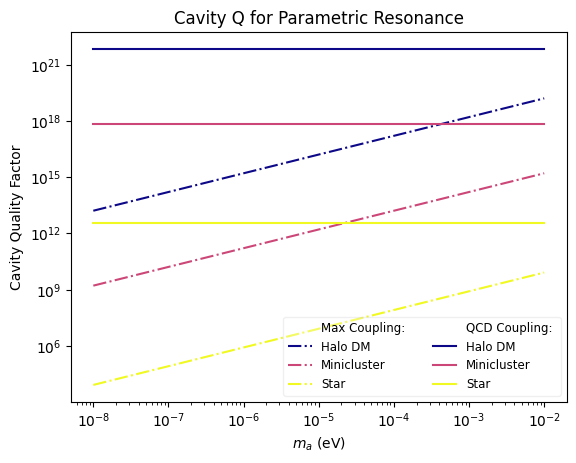}
\caption{Cavity resonant quality $Q$ required for parametric resonance in the case of QCD and maximal couplings for structures of varying density.}
\label{fig:quality_plot}
\end{figure}

This condition for unstable parametric resonance, interpreted as a mininum cavity $Q$, is plotted in Fig.~\ref{fig:quality_plot} for both the maximum allowed coupling for axion like particles allowed by astrophysics, and the expected KSVZ coupling for the QCD axion, and for densities expected for a smooth dark matter background, an intermediate axion minicluster, and a dense axion star as described above, over a range of haloscope-accessible axion masses.  Evidently, the simplest models of smooth dark matter and a QCD axion are out of reach from existing experimental capabilities by 10 orders of magnitude.  Interestingly, however, were the Earth located within an axion star, and the axion mass known, it would be technically possible to build a resonator that would induce a strong signal through unstable parametric resonance.

\section{Conclusion}

We conclude that, while there exist extreme models of axion dark matter structure and unexcluded axion-like-particle couplings for which parametric resonance could be induced with existing resonator technology, the technique is not directly applicable to improve axion dark matter searches.  What is unusual about this process is that it produces an exceptionally large amount of power given how feeble the coupling is.  In the far far future, perhaps it may be useful to extract energy from axion stars.  However, future work involving more sophisticated nonlinear couplings, the development of higher Q resonators, or revelations about dark matter substructure may increase the relevance of this approach.

\section{Acknowledgments}
This work was supported by
the U.S. Department of Energy through Grants No. DE-
SC0011665 and the NSF REU Program through award No. 2243362.

\bibliography{main}
\end{document}